# Initial Findings from a Study of Best Practices and Models for Cyberinfrastructure Software Sustainability

Craig A. Stewart, Julie Wernert, Eric A. Wernert, William K. Barnett, Von Welch
Pervasive Technology Institute
Indiana University

stewart@iu.edu, jwernert@iu.edu, ewernert@iu.edu, barnettw@iu.edu, vwelch@indiana.edu

**Abstract**

We present a set of common themes and recommendations extracted from in-depth interviews with the leaders of 12 distinct cyberinfrastructure software projects. These interviews were conducted as part of a larger study to identify and elucidate the best practices and management models that lead to sustainability for cyberinfrastructure software. Respondents in a formal survey of cyberinfrastructure users identified these projects as good examples of sustained software initiatives. While there is clearly no single method or plan that will guarantee sustainability for all projects, we can draw general guidance from these exemplars. This paper presents the common themes, ideas, and recommendations that emerged from those interviews.

## 1. Motivation

Software is central to the NSF's vision of a Cyberinfrastructure Framework for 21st Century Science and Engineering (CIF21). Through multiple reports (e.g. *A Vision and Strategy for Software for Science, Engineering, and Education*) and major solicitations (e.g., Software Infrastructure for Sustained Innovation), it is clear that software is essential to computational and data-enabled science. With this fundamental statement as the cornerstone of our work – and building upon the recommendations resulting from the 2009 NSF-funded Workshop on Cyberinfrastructure Software Sustainability and Reusability – our study of cyberinfrastructure projects has aimed to identify the operational software management and deployment practices, organizational principles, administrative structures, and governance models that lead to software projects that are well sustained from the standpoint of creators and users, and which can be counted on to provide a robust and sustained software tools to enable a comprehensive implementation of NSF's CIF21 vision. We have identified attributes leading to scientific impact and innovation, as well as cultural norms that may impede impact, growth and sustainability, and have developed a set of finding and recommendations to guide the funding, development, deployment, and operation of future cyberinfrastructure software initiatives.

## 2. Methodology

The study was comprised of two major activities: a formal survey of software producers and users, and a set of detailed case study interviews. The survey was sent to the principal investigators on every NSF grant over the preceding five years – over 5,000 PIs covering all NSF directorates. The survey asked a range of questions on practices that aid or hinder software sustainability. We received 811 valid responses. (A report of the complete results of that survey is forthcoming.)

Among the questions on the survey were ones that asked respondents to identify software tools that they considered successfully sustained. We used those responses to identify potential software projects for in-depth case studies, adjusting the list for those that were not specifically focused on cyberinfrastructure (e.g., text editors and operating systems.) The final candidate projects were selected to represent a broad cross-section of scientific domains and software levels (utilities, middleware, end-user applications), and thus, did not fully represent the most frequently cited projects. A total of 12 case study interviews were conducted, usually with the project director or one or more lead developers, using a standard set of questions. Interviews were conducted in-person whenever possible, but by phone as necessary.

## 3. Preliminary Findings from Case Studies

Through this study, we discovered that there are a myriad of reasons that software projects fail to reach sustainability; the reasons for failure are often unique, and have little in common with the failures experienced in other projects. Conversely, we have discovered that initiatives that do succeed share many commonalities; that is, successful projects have many similar fundamental traits, regardless of the type of software, the community it supports, or its licensing model. Further, by employing methods and tools traditionally found in social science research (e.g., surveys, interviews), we gained knowledge about what the scientific community expects from a software project, insights into traditional and alternative methods funding of software initiatives, measures for gauging impact, and ideas about how software might "grow old gracefully" as it continues to serve its core audience. The following sections summarize the common themes, ideas, and recommendations that emerged from the case study interviews.

### 3.1. The Big Picture

- **The general landscape.** The concept of openness is broadly and deeply viewed as a cornerstone of scientific discovery. As a matter of principle, there is near-universal agreement that scientific software should be free, open, reliable, flexible, regularly updated and expanded, well supported and documented, and easily and continuously accessible. As a matter of practice, however, there is little agreement about how to accomplish these ideals, what standards and measures are necessary for evaluation and adoption, or which financial models will most successfully move an initiative toward long-term sustainability. Further, at the present time, there is little motivation or incentive to depart from the standard assumption that ongoing funding from the National Science Foundation or other federal agencies is the only path to sustainability.

- **Licensing milestone.** The licensing decision is a crucial milestone in all software projects. This decision will impact the project throughout the course of its lifetime. Many principals assert that while universities are often hesitant to accept the notion of "open source" intellectual property, the most permissive license possible will reduce overhead, promote adoption, and lend to the project's long-term sustainability.

- **The greater good.** Whether it be solving grand challenges or simply decreasing the amount of time it takes to produce new science, long-term sustainability is aided when an initiative serves a common or greater good that benefits society as a whole.

### 3.2. Leadership and Management

- **Strong, visionary leadership**. Strong, committed, visionary leadership is central to the development of sustained software initiatives. Principals in software initiatives that are sustained are recognized by their peers as thought leaders within their respective fields and are visible and proactive in engaging their user communities and promoting their initiatives. Further, in most successful initiatives, there is a solid, trusted cohort of operational leadership that supports the vision set forth by the principal investigators.

- **Centralized leadership**. Centralizing the leadership within a small, core group of principals with clearly defined roles increases the long-term viability of a software initiative. Broadly dispersed, multi-institutional initiatives often suffer from layers of institutional bureaucracy, duplication of effort, and miscommunication.

- **Business-oriented approaches**. Open-source software initiatives share many common challenges with small, non-profit businesses and must be managed as such to reach and maintain a level of sustainability. Like any successful business, attention must be paid to branding, product portfolio, customer service and support, quality control, reputation and risk management, industry trends, stakeholders, staffing, budgets, long-term growth, and sustainability. That said, the vast number of principal investigators and researchers in the scientific domain lack the training, expertise, awareness, time, and, in some cases, inclination or aptitude to become functionally proficient in these areas crucial to sustainability. A small-business boot camp designed around the tolerances and motivations of scientists and researchers should be developed to, at a minimum, provide an overview of these areas and proactively create an awareness of the key components required for sustainability.

### 3.3. Funding

- **Federal lifecycle funding for critical software**. Investments in "critical-path" scientific software – the utilities, end-user applications, and/or middleware essential to research, discovery, and innovation – is a key component of the national scientific cyberinfrastructure, and in many cases, has a life cycle many times greater than similar investments in hardware, and therefore should be viewed and funded as "infrastructure," equally critical to the process of research, discovery, and innovation.

- **Alternate funding for less critical tools**. While some software initiatives are mission critical to the broader US scientific enterprise, others – while still very important – are not. In an era of ever-shrinking budgets, principal investigators should be encouraged (and in some cases required) to explore non-Federal options for ongoing expansion and maintenance. A paradigm shift in the widely held premise that "either the software must be federally-funded or it will have to die," is shortsighted and does not protect the vast investments previously made. (Alternately stated, "I'll do this as long as the government pays me to do it, otherwise its someone else's concern.")

  **Third party measurement of impacts.** The National Science Foundation and other funding agencies should consider gauging the impacts of a project through sources other than (or in addition to) the principal investigators. Principal investigators are often very "hands off" and unable to accurately measure impacts. Further, they are often unaware of the broader research landscape (beyond their own domain), and thus funding agencies often "reinvent the wheel" in funding projects that have already been funded in a another domain, program and/or agency.

### 3.4. Developers and Users

- **Developers engaged with users.** A centralized development team that has regular and routine contact with users and leadership is essential to sustainability. Developers in sustained initiatives are integrated into the support mechanisms of their initiatives, regularly engaging users by responding to help desk inquiries, monitoring and participating in listserv, wiki, or blog discussions, and presenting workshops or training classes.

- **Domain experts engaged with developers.** Software initiatives that pair subject matter (domain) expertise with technological and engineering expertise have an increased probability of broad adoption and sustainability. Such initiatives can adapt to changing needs, technologies, and trends within a specific domain, while building robust yet flexible software that can be easily expanded or modified, and possibly adopted outside its original project or domain.

### 3.5. Planning for Sustainability

- **Detailed sustainability plans for ongoing projects.** A phased sustainability management plan, ranging from preliminary to advanced, should be part of any proposal that requests funds for development, expansion, or ongoing maintenance of scientific software. While successful software initiatives have different needs, will follow different timelines, and will reach varying numbers and sizes of communities, most follow a small number of proven trajectories that achieve predictable milestones. Thus, critical planning as to long-term viability can and should be required in all stages of funding and development, especially at the point where second- and third-round funding is being sought. Principal investigators should be expected to provide increasingly finer levels of granularity with their proposals.

- **Objective sustainability assessments for new projects**. At the same time, not all software is intended, nor does it need, to be sustained. Further, some software projects are unnecessary, as they duplicate previous or current efforts and/or existing commercial or open-source solutions. An instrument (e.g., in-depth survey, assessment, etc.) is needed to help objectively address sustainability and life cycle factors (e.g., expected number of users/market share, new and unique capabilities of the proposed software, funding trajectory and contingencies, commercial/pre-existing solutions, etc.) as a prerequisite to early-stage funding for new initiatives.

- **Aggregation for economies of scale.** Referencing a proven model that has evolved over decades in the atmospheric sciences (i.e., UCAR/Unidata), an aggregation of multiple software initiatives in a given domain should be considered when logistically and financially feasible. Centralizing the software activities for specific fields of study would, over time, create a critical mass of thought leadership, as well as technical and domain-specific expertise. This would also aid in eliminating redundancies and duplication of effort, while allowing for economies of scale to be achieved in support, training, documentation, release, archiving, and administrative/management activities.

**3.6 Software Sustainability Institute**

A software sustainability institute should be created and vested with the authority to guide and oversee the national investment in software infrastructure. Functions and services housed within such an institute might include:

- Development and oversight of standardized quality control, testing, regression, and documentation processes. There is little consistency across projects relative to these processes, and virtually no agreed upon standards by which to evaluate projects for critical-path adoption and/or additional funding.

- A central, common repository for all nationally funded software. This repository should address not only long-term access, but also the loss of intellectual property associated with student matriculation, investigator retirement/death, a change in research focus, or benign neglect/abandonment.

- A software orphanage/retirement center – a software escrow, of sorts – where software projects nearing the end of their life cycle but still viable and actively used by members of the scientific community, or those for whatever reasons that are no longer in contention for additional funding, may be housed to ensure basic maintenance and accessibility until such time that there is no longer significant demand.

- A central resource for services and consulting where smaller projects, or those with specific, short-term needs, could acquire expert consultation or enterprise services on an as-needed basis via a charge-back model. The institute would provide a basic set of in-house services and also serve as a broker, helping to facilitate the pairing of projects in need of certain competencies or resources with known expertise and/or reliable service providers.  (For example, the institute might help a smaller project needing to offer user support or knowledge base functionality to acquire these services from a larger project with a robust, extensive support mechanism already in place.) Longer-term consultations or ongoing service-provider agreements could be facilitated through future grant subcontracts.) Not only would this generate economies of scale and decrease duplication of efforts, it would allow "consumers" of these resources and services to easily procure from trusted, vetted sources with proven track records, as well as serve to break down the silos that exist within scientific communities

- A "think tank," of sorts, where the issues and requirements for sustainability can be debated and, ultimately, authoritatively determined outside of the funding context.

**Acknowledgements**

The authors would like to thank members of the scientific and higher-education software development communities for contributing their expertise to this research and participating in extended discussions about their software initiatives. This work is made possible by the EArly-concept Grants for Exploratory Research (EAGER) Program funded by the National Science Foundation through the award: OCI-1147606.